\documentclass[a4paper]{jpconf}
\usepackage{graphicx}
\begin{document}
\title{Data Preservation and Long Term Analysis in High Energy Physics}

\author{D M South, on behalf of the DPHEP Study Group}

\address{Deutsches Elektronen Synchrotron, Notkestra\ss e 85, 22607 Hamburg, Germany}

\ead{david.south@desy.de}

\begin{abstract}
Several important and unique experimental high-energy physics programmes
at a variety of facilities are coming to an end, including those at HERA, the B-factories
and the Tevatron.
The wealth of physics data from these experiments is the result of a significant financial 
and human effort, and yet until recently no coherent strategy existed for data preservation
and re-use. 
To address this issue, an inter-experimental study group on data preservation and
long-term analysis in high-energy physics was convened at the end of 2008, publishing an
interim report in 2009.
The membership of the study group has since expanded, including the addition of the LHC
experiments, and a full status report has now been released.
This report greatly expands on the ideas contained in the original publication and provides
a more solid set of recommendations, not only concerning data preservation and its
implementation in high-energy physics, but also the future direction and organisational
model of the study group.
The main messages of the status report were presented for the first time at the 2012
International Conference on Computing in High Energy and Nuclear Physics and are
summarised in these proceedings.
\end{abstract}

\section{Introduction}
\label{sec:intro}

The last $60$~years have produced a wealth of high-energy physics (HEP) results from a variety
of often very different experiments.
Over this period, new energy and intensity frontiers have been continuously probed, using
increasingly complex accelerator installations.
A growth in the size of the necessary international collaborations associated to
the experiments may also be observed, as well as an increase in the diversity of the
data management.
The culmination of this period is not only the Large Hadron Collider at CERN, whose
proton-proton collisions at a record centre of mass energy of $8$~TeV will be the focus
of this year's summer conferences, but also a legacy of results from many other
experiments who have only recently stopped data taking and are at this time in the
process of publishing their final measurements.
Furthermore, it is hoped that several other projects such as the Super-B factories,
the ILC and the LHeC will complement the LHC physics programme in the next period.


A central question which has up to now not been properly addressed in the field of HEP
is what to do with the data after the collisions have stopped and the planned analysis
programme of an experiment has been completed.
Until recently, there was no clear policy on this and it is likely that older HEP experiments
have simply lost the data.
Data preservation, including long term access, is generally not part of the planning,
software design or budget of an experiment.
Previous HEP data preservation initiatives have in the main not been planned by the original
collaborations, but have rather been the effort a few knowledgeable people, typically
ex-collaborators, with a keen interest in preventing the data disappearing forever.


So why is it difficult to preserve HEP data, in particular seeing as a diverse range of other
scientific disciplines such as astrophysics, life sciences and molecular biology have to some
level integrated successful data preservation programmes into their respective fields?
The scale and distribution of HEP data complicates the issue, where the question of
custodianship, and who is ultimately is responsible for the data  - the experiments or
the host computing centres - may also be unclear.
Aging, unreliable hardware and highly specialised experiment-specific software
contribute to the problem.
Key resources, both funding and person-power expertise, tend to decrease once the data taking
at a HEP experiment stops, when the focus tends to shift elsewhere.
And perhaps the most important ingredient is to examine if there is a physics case to
answer: can the benefits be balanced against the necessary cost and effort in developing an
effective and robust data preservation programme concerning HEP data?

\section{The DPHEP study group}

\begin{figure}[h]
  \begin{center}
    \includegraphics[width=\textwidth, clip=true, trim=4cm 5cm 0cm 6cm]{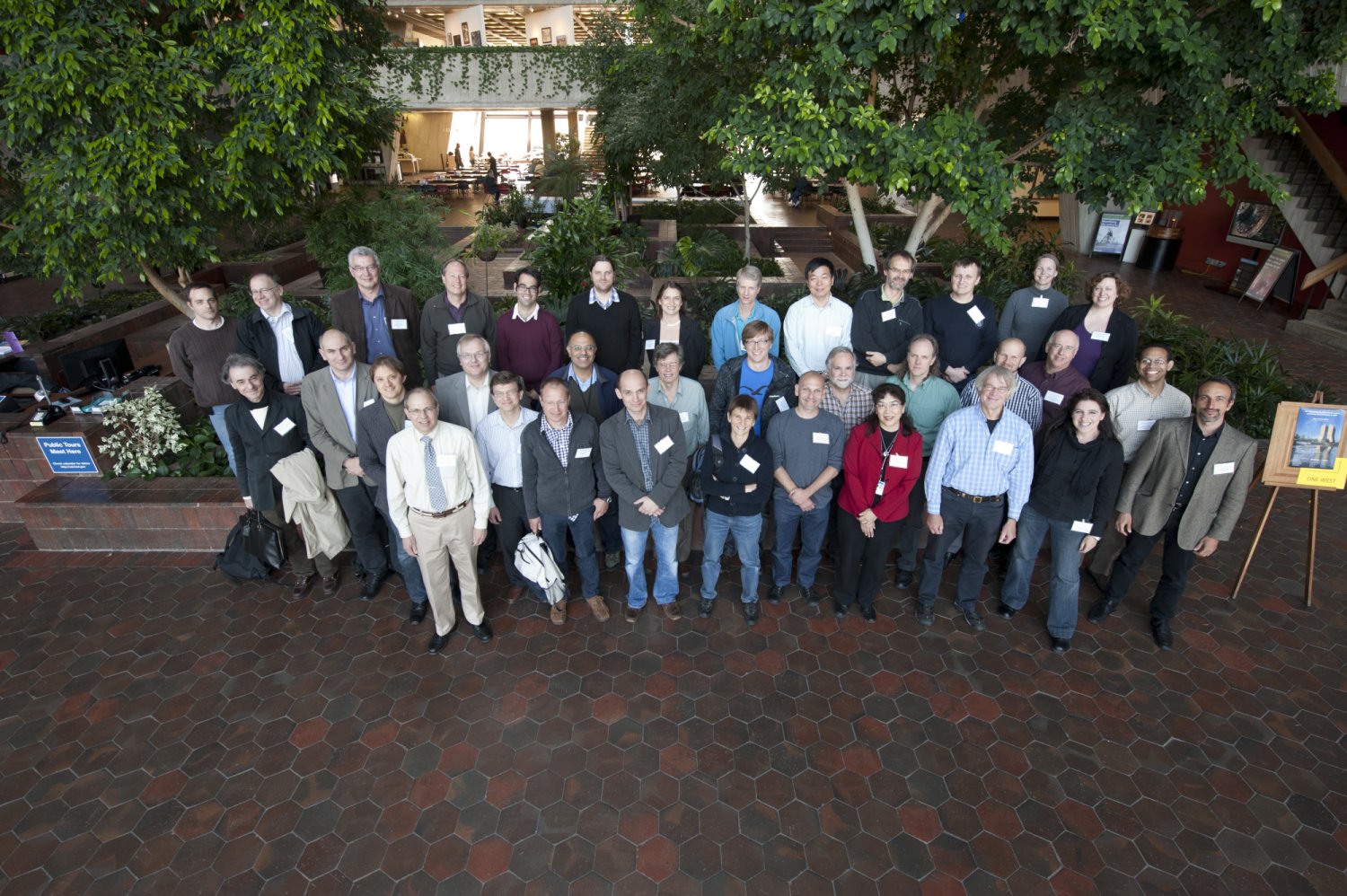}
  \end{center}
  \vspace{-0.3cm}
 \caption{\label{fig:people}Participants of the fifth DPHEP workshop at Fermilab, May 2011.}
 \vspace{-0.3cm}
\end{figure}

An international study group on data preservation and long term analysis in high-energy
physics, DPHEP~\cite{dphep}, was formed at the end of 2008 to address this issue in a systematic way.
The aims of the study group include to confront the data models, clarify the concepts, set a 
common language, investigate the technical aspects, and to compare with other fields 
such as astrophysics and those handling large data sets.
The experiments BaBar, Belle, BES-III, CLAS, CLEO, CDF, D{\O}, H1, HERMES and ZEUS are
represented in  DPHEP; the LHC experiments ALICE, ATLAS, CMS and LHCb joined the study
group in 2011.
The associated computing centres at CERN (Switzerland/France), DESY (Germany),
Fermilab (USA), IHEP (China), JLAB (USA), KEK (Japan) and SLAC (USA) are all also
represented in DPHEP.  
 

A series of five workshops have taken place over the last three years, and since 2009
DPHEP is officially endorsed with a mandate by the International Committee for Future
Accelerators, ICFA.
The initial findings of the study group were summarised in a short interim
report in December 2009~\cite{dpheppub1} and a full status report was released in
May 2012~\cite{dpheppub2}, to coincide with the International Conference on Computing
in High Energy and Nuclear Physics (CHEP) 2012~\cite{chep}.
A parallel DPHEP meeting was also hosted by the conference~\cite{dphepatchep}.
The report contains: a tour of data preservation activities in other fields; an expanded
description of the physics case; a guide to defining and establishing data preservation
principles; updates from the experiments and joint projects, as well as person-power
estimates for these and future projects; the proposed next steps to fully establish
DPHEP in the field.


Only a brief summary of the full status report is presented in these proceedings, highlighting
the areas presented in the plenary session at the CHEP 2012 conference.
Additional data preservation contributions at CHEP 2012 from the BaBar, H1,
HERMES and ZEUS experiments, as well as the DESY-IT division, may be also be found
in the conference proceedings or on the conference webpages~\cite{chep}.

\section{The physics case for data preservation}

When building the physics case for data preservation, four main categories have
been identified:
Firstly, data preservation is beneficial to the long-term completion and extension
of the physics programme of an experiment.
In the case of the LEP experiments a considerable tail exists in the
publication rate, which continues today, and a similar trend is predicted by the HERA
experiments and BaBar.
Up to $10$\% of papers are finalised in the post-collisions period, and prolonging
the availability of the data may result in a gain in scientific output of an experiment.


Secondly, cross-collaboration and combination of data from multiple experiments may
provide new scientific results, with improved precision and increased sensitivity.
This may occur during the active lifetime of similar experiments at one facility, such as
those at LEP, HERA, or the Tevatron, but may also occur later across larger boundaries,
such as combinations of Belle and BaBar or Tevatron and LHC data.
The preservation of such data would facilitate the comparison of complementary
physics results as well as allowing the independent verification of experimental
observations.


Thirdly, it may be useful to revisit old measurements or perform new ones
with older data.
Access to newly developed analysis techniques as well as the possibility to perform
comparisons to state-of-the-art theoretical models may produce improved or even
new physics results.
Furthermore, unique data sets are available in terms of initial state particles or centre
of mass energy or both, such as the PETRA $e^+e^-$, HERA $e^{\pm}p$ and Tevatron
$p\bar{p}$ collision data, as well as data from a variety of fixed target experiments.
More recently, the early LHC data, taken at centre of masses of $900$~GeV and $2.36$~TeV,
as well as the low pile-up $7$~TeV data taken in 2010 also provide unique opportunities.


Finally, the value of using real HEP data for scientific training, education and outreach
cannot be understated.
Providing a wide variety of HEP data sets for such analysis, with a corresponding wide
variety of associated exercises and teaching programmes, is clearly beneficial in 
attracting a new generation of inquisitive minds to the field.


A more detailed description of the physics case for data preservation can be found
in the DPHEP status report~\cite{dpheppub2}, which includes specific examples of
analyses using older data from among others LEP and PETRA, in addition to a description
of the potential of the data from the experiments in the final analysis phase: the
B-factories, HERA and the Tevatron.

\section{Models of data preservation in high-energy physics}

In order to develop a solid definition of models of data preservation, it is first important
to ask the question: what is HEP data?
The digital information, the data themselves, are clearly crucial, but at least for the
pre-LHC experiments volume estimates for preservation are of the order of a few to
$10$~PB, which is certainly within the storage capabilities of today's HEP based
computing centres.
The range in data volume to be preserved is often a result not only of different sized
data sets, but different types of data: from the basic level raw data, through reconstructed
data, up to the analysis level ntuples.


However, if one thing may be learned from previous enterprises, it is that the conservation
of tapes is not equivalent to data preservation, and that providing not only the hardware to
access the data but also the software and environment to understand the data are the
necessary and more challenging aspects.
Therefore, in addition to the data the various software, such as simulation, reconstruction
and analysis software need to be considered.
If the experimental software is not available the possibility to study new observables or to
incorporate new reconstruction algorithms, detector simulations or event generators is lost.
Without a well defined and understood software environment the scientific potential of the data
may be limited.


Just as important are the various types of documentation, covering all facets of an experiment.
This includes the scientific publications in journals and online databases such as
INSPIRE~\cite{inspire} and arXiv, published masters, diploma and Ph.D. theses, as well as a
myriad of internal documentation in manuals, internal notes, slides, wikis, news-groups and so on.
Detailed information about analyses may only be available in internal notes, which may not be
easily available, electronically or otherwise.
Many types of internal meta-data may also exist, such as the details of the detector layout and
performance, hardware replacements, manuals or the documentation of meetings. 


Finally, the often unique expertise and contributions of collaboration members must also be
considered as another component of HEP data.
Particular care is needed to ensure crucial know-how does not disappear with losses in person-power,
which is liable to happen towards the end of an experiment as described in section~\ref{sec:intro}.


Considering this inclusive definition of HEP data, a series of data preservation levels are established
by the DPHEP study group, as summarised in table~\ref{tab:levels}.
The levels are organised in order of increasing benefit, which comes with increasing complexity
and cost.
Each level is associated with use cases, and the preservation model adopted by an experiment should 
reflect the level of analysis expected to be available in the future.
These are the original definitions of DPHEP preservation levels from the interim report~\cite{dpheppub1},
which remain valid, although the interaction between the levels is now better understood: whereas the original
idea was a progressively inclusive level structure, the levels are now rather seen as complementary initiatives.
The four levels represent three different areas: documentation (level 1), outreach and simplified formats
(level 2) and technical preservation projects (levels 3 and 4).

More generally, the implementation of a data preservation model as early as possible during the
lifetime of an experiment may greatly increase the chance that the data will be available in the
long term, and may also simplify the data analysis in the final years of the collaboration. 
Planning a transition of the collaboration structure to something more suited to a long-term
organisation also makes it easier to address issues such as authorship, supervision and access.


\section{Current data preservation projects}

Some examples of ongoing projects at the experiments and laboratories involved in the DPHEP study group
are described in the following, placing them in the context of the data preservation levels described in the
previous section.
More details on each of the preservation levels, in addition to a full description of the various individual and
joint data preservation initiatives can be found in the DPHEP status report~\cite{dpheppub2}.

\subsection{Level 1 projects: documentation}

Although a level 1 preservation model, to provide additional documentation, is considered 
the simplest, this still requires some often substantial activity by the experiment.
The organisation of documentation requires a significant effort, with much material from
pre-web days or using a range of often older web applications, and dedicated task forces
have been set up by many of the experiments to perform this task.
The organisation and cataloguing of non-digital documentation and the scanning or
photographing of materials such as papers and talks, notes, drawings, detector
schematics, blueprints, logbooks is in itself a large scale project.
New ``virtual archives'' have been established by the experiments for newly digitised
material, as well securing dedicated new locations to store the newly sorted paper items. 


\begin{table}[t]
  \begin{center}
    \caption{\label{tab:levels}Data preservation levels defined by the DPHEP study group~\cite{dpheppub2}.}
    \centering
    \begin{tabular}{{c}{l}{l}}
      \br
      Level & Preservation Model & Use Case\\
      \br
      1 & Provide additional documentation & Publication related info search \\
      \mr
      2 & Preserve the data in a simplified format & Outreach, simple training analyses\\
      \mr
      3 & Preserve the analysis level software and & Full scientific analyses,\\
      & data format & based on the existing reconstruction\\
      \mr
      4 & Preserve the simulation and reconstruction & Retain the full potential of the\\
      & software as well as basic level data & experimental data\\
      \br
    \end{tabular}
  \end{center}
\end{table}

Digital documentation, which encompasses online shift tools, detector configuration files,
electronic logbooks and various webpages of meetings, talks, presents its own challenges.  
The HERA experiments have replaced old, dedicated web-servers by virtual machines (VMs),
hosted by the computer centres, thus relieving the burden of updating the hardware and
migrating the data in the future.
The migration of old webpages to newer technologies such as wikis is also proving
to be beneficial.


The use of external services for hosting collaboration material has also been examined.
The internal notes from all HERA collaborations are now available on INSPIRE, so that
once again the experiments no longer need to provide dedicated hardware.
These INSPIRE records, which can now be linked to other records, remain password
protected for now, but it is now simple to make them publicly available in the future.
The ingestion of other collaboration documents by INSPIRE is under discussion,
including theses, preliminary results, conference talks, proceedings.
The BaBar, CDF and D{\O} and experiments are also working on similar projects with INSPIRE.

\subsection{Level 2: Simplified formats and outreach}

A level 2 preservation effort, to conserve the experimental data in a simplified format, is considered to
be unsuitable for high level analyses, lacking the depth to allow, for example, detailed systematic
studies to be performed.
However, such a format is ideal for education and outreach purposes, which many experiments in the
study group are also actively interested in, and may also be useful to test new theoretical models as
an easy way to share complicated data.


Within DPHEP there are generic ideas, such as common formats and user interfaces.
Such formats in HEP are typically based on ROOT~\cite{root}, containing particle
4-vectors and simple event information, providing an input to simple analyses requiring
composite-particle reconstruction or to those looking for signals in the data.
Outreach initiatives are in place at BaBar, Belle and the LHC experiments and a
multi-experimental project is clearly desirable, coordinated via DPHEP, and based in several
locations such as CERN, FNAL and DESY, with associated tutorials linked to preserved HEP
data from several sources.

\subsection{Levels 3 and 4: Technical projects}

The main focus of the data preservation effort are the technical projects to preserve
not only the data, but also to ensure long term access and usability.
Whereas level 3 provides access to analysis level data, MC and the analysis level software, level 4
additionally includes access to the reconstruction and simulation software.
Past experiences with old HEP data indicate that new analyses and complete re-analyses are 
only possible if all the necessary ingredients to retrieve, reconstruct and understand the data are 
accounted for.
Only with the full flexibility does the full potential of the data remain, equivalent to level 4 preservation.
Accordingly, the majority of participating experiments in the study group plan for a level 4 preservation
programme, although different approaches are employed concerning how this goal should be achieved. 
Typically, this can be realised in two ways: either keep the current environment alive as long as possible,
or adapt and validate the environment to future changes as they happen.
These two complementary approaches are taken at SLAC and DESY respectively, both employing
virtualisation techniques, but in rather different ways.

\begin{figure}[h]
  \begin{center}
    \includegraphics[width=0.85\textwidth]{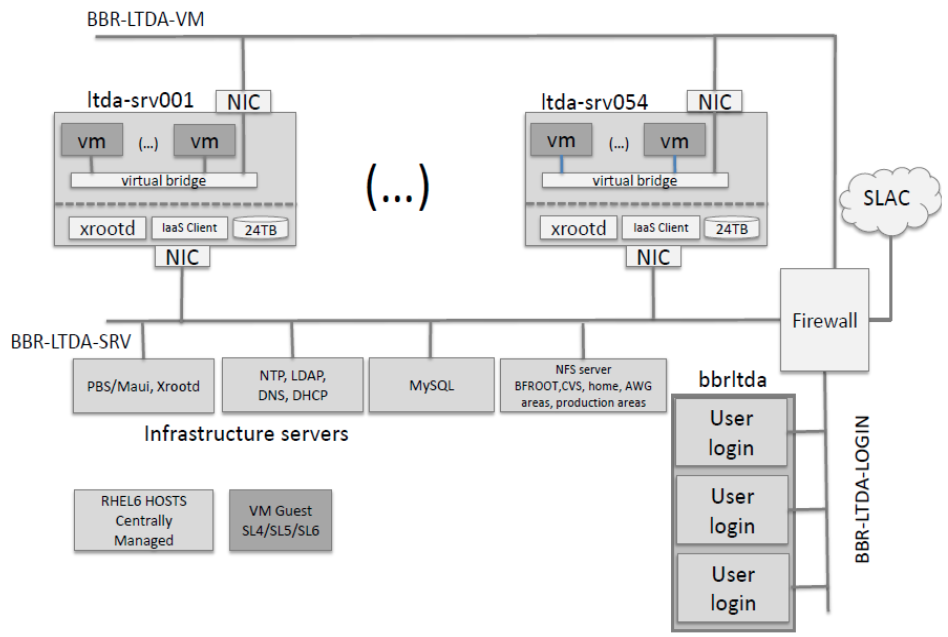}
  \end{center}
\vspace{0.2cm}
  \begin{center}    
    \includegraphics[width=0.95\textwidth]{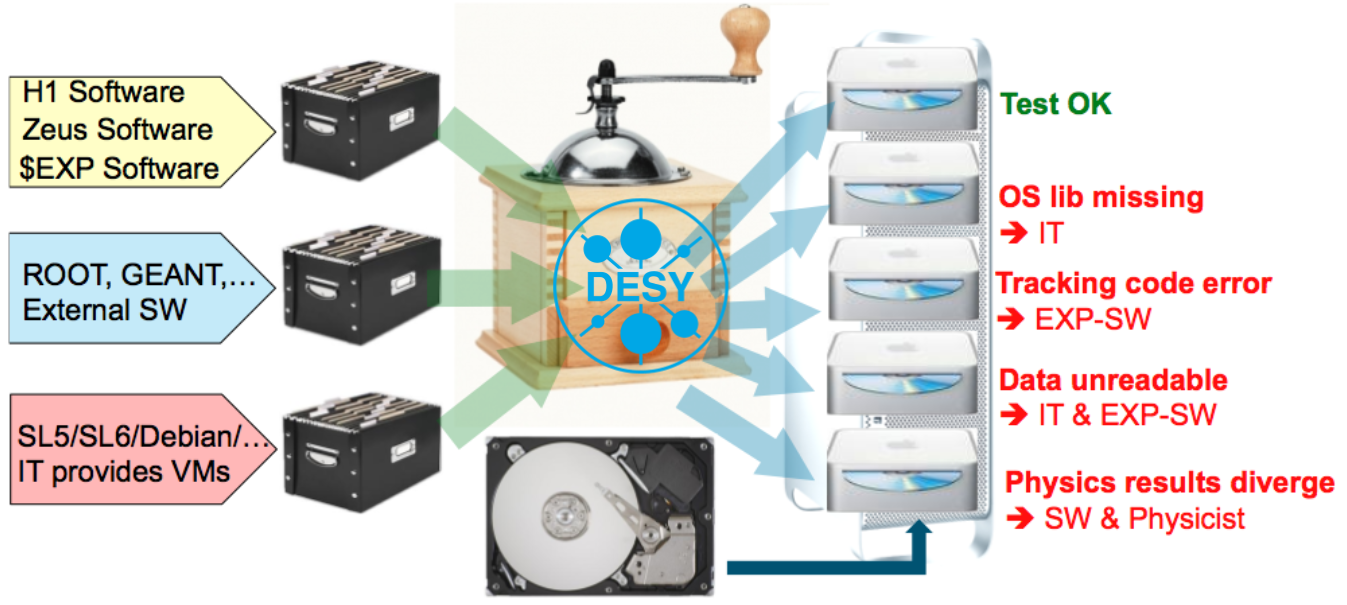}
    \caption{\label{fig:techs}Top: The networking schematic of the BaBar LTDA archival
      system employed at SLAC. Bottom: An illustration of the generic validation framework at DESY.}
 \end{center}
\end{figure}


The BaBar Long Term Data Access (LTDA) archival system is now installed for analysis until at
least 2018.
The deployment of the LTDA required a large scale investment, with over 70 new dedicated servers,
where crucially the resources for the project were taken into account in the BaBar funding model during
the analysis phase. 
Isolated from the rest of SLAC computing, it uses virtualisation techniques to preserve an existing,
stable and validated platform and provides a complete data storage and user environment in one system.
A crucial part of the design is to allow frozen, older platforms to run in a secure computing 
environment.
It is now known that a na\"{i}ve virtualisation strategy is not enough, and that an old operating system (OS)
cannot simply be supported forever, when the security of system may come under threat.
The LTDA safeguards against this by having a clear network separation via firewalls of the part
storing the data (running a more modern OS) and the part for performing analysis (the desired older OS),
as illustrated in figure~\ref{fig:techs} (top).
More than 20 analyses are now using the LTDA system, where from the user's perspective it appears
very similar to the classical BaBar infrastructure.


The alternative approach taken at DESY is to employ an automated validation system to
facilitate future software and OS transitions.
In this system, virtualisation is used for flexibility, in order to host different configurations
of a software environment in one coherent set up.
This is illustrated in figure~\ref{fig:techs} (bottom), where the separate inputs of
experimental software, external software and OS are combined to form a VM, which is then
deployed on many systems where a series of experimental tests are performed. 
A successfully validated environment recipe can then be installed on a future resource, such
as the Grid or an IT-cluster, to provide a working environment for the experiment.
An essential component of such a project is a robust definition of a complete set of
experimental tests.
The nature and number of the tests is dependent on desired preservation level;
both H1 and HERMES aim for full level 4 preservation, ZEUS between levels 3 and 4.
Providing a display of the validation results in a comprehensible way is also required,
which is fulfilled using a web-based interface.
The characteristics of each test determines the nature of the result, which could be simple
yes/no, a set of plots, ROOT files, a text-file containing keywords or of a certain length
and so on.
The common OS baseline of SLD5/32-bit was achieved in 2011 by all of the HERA
experiments and the ongoing validation of 64-bit systems is the next major objective 
towards future OS migrations.
Additionally, the system has already been useful in detecting problems that were
visible only in newer software.

\section{Conclusion and future working directions}

The DPHEP study group represents the first large scale effort to address data preservation in the
field of high-energy physics.
The initial make up of the group was driven by the coincidence of the end of data taking at several
large colliders, but has grown to include others including the LHC experiments.
The activity of the group over the last three years has led to an increased understanding of the relevant
issues, enabling problems to be addressed, recommendations to be formulated and multi-experiment
projects to begin.
The recent DPHEP publication, which contains a comprehensive appraisal of data preservation in HEP,
represents a significant milestone and concludes the initial period.
To gain the most benefit from the work done so far, a transition from the current study group structure
to a new, full time organisation is required.

\begin{figure}[]
  \begin{center}
    \includegraphics[width=0.7\textwidth]{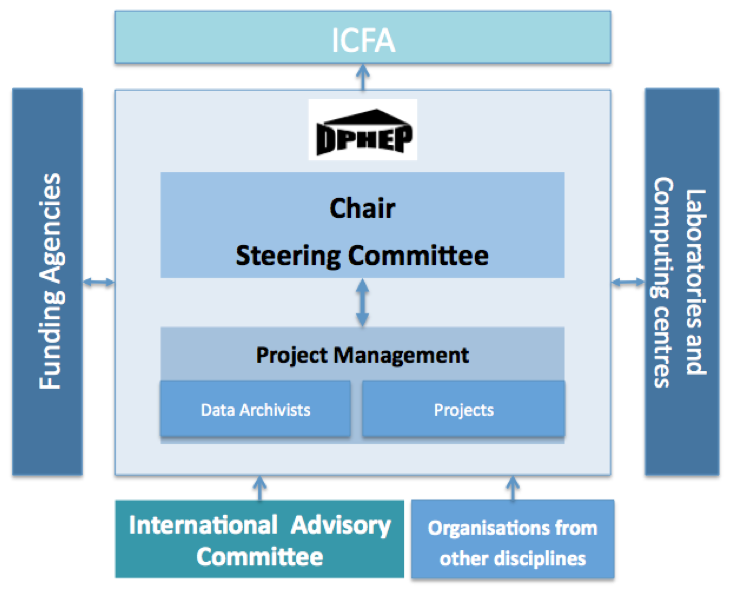}
  \end{center}
 \caption{\label{fig:dpheporg}The proposed DPHEP organisation and its associations.}
\end{figure}

The DPHEP organisation should retain the basic structure of the study group, with links to the host
experiments, laboratories, funding agencies and ICFA.
The main difference is the installation of a dedicated full time position, the DPHEP Project Manager,
who acts as the main day-to-day operational coordinator, as illustrated in figure~\ref{fig:dpheporg}.
The organisation will nevertheless continue to investigate and take action in areas of coordination,
preservation standards and technologies, as well as expanding the experimental reach and
inter-disciplinary cooperation.
The next DPHEP workshop, which will focus on this last point, will take place in Autumn 2012.

\section*{References}

\begin{thebibliography}{50}

\bibitem{dphep}
  Study group on Data Preservation and Long Term Analysis in HEP, DPHEP; http://dphep.org

\bibitem{dpheppub1}
 Asner D et al. (DPHEP Study Group) 2009
 {\it Preprint} arXiv:0912.0255

\bibitem{dpheppub2}
 Akopov Z et al. (DPHEP Study Group) 2012
 {\it Preprint} arXiv:1205.4667

\bibitem{chep}
 International Conference on Computing in High Energy and Nuclear Physics, CHEP 2012 (New York City); http://www.chep2012.org

\bibitem{dphepatchep}
 DPHEP@CHEP2012; https://indico.cern.ch/conferenceDisplay.py?confId=171962

\bibitem{inspire}
  INSPIRE is the successor to the popular SPIRES HEP database; http://inspirehep.net

\bibitem{root}
  Brun R and Rademakers F {\it Nucl. Instrum. Meth.} A {\bf 389} (1997) 81


\end{thebibliography}
\end{document}